# TURNING THE DISRUPTIVE POWER OF BLOCKCHAIN IN THE INSURANCE MARKET INTO INNOVATIVE OPPORTUNITIES

Wadnerson Boileau

University of South Florida (USF), United States of America

*ABSTRACT*

*Insurance has been around for more than centuries. This risk mitigation strategy has been utilized in maritime commerce as early thousand years ago, where Asian merchant seafarers were pooling together their wares in collective funds to pay for damages of individual's capsized ship. In 2018, insurance industry made up 6% of global domestic product, and amounted to about 7-9% of the U.S.GDP;2020, the industry net premiums totalled $1.28 trillion, by 2030, blockchain insurance market value is estimated to reach $39.5 Billion. Despite of growing reform, the insurance market is dominated by intermediaries assisting people to match their insurance needs. While many predictions focused on artificial intelligence, cloud computing, blockchain stands out as the most disruptive technology that can change the driving forces underlying the global economy. This paper presents a blockchain business use case and how insurance market can turn disruptive power of this technology into innovative opportunities.*

## 1. INTRODUCTION

Blockchain in recent years becomes one of most powerful technologies in the academic sector and financial industry. This technology is constantly progressing to prove its potentiality to digitally transforming processes inside and outside organizations and influencing coordination as a mechanism to rule companies. Financial sector in many instances relies on blockchain to find durable solutions to some of the common problems faced in the insurance industry including lower frictional cost of frivolous claims, increasing efficiency and service delivery through technological innovations, and growing the customer base through trust. The paper aims to provide a clear-eyed view of how blockchain can be used in the insurance industry for process innovation. Throughout this paper, we focus on business blockchain use case showing where the technology can bring efficiencies which enables the sector to turn the disruptive power of blockchain into opportunities.

## 2. LITERATURE REVIEW

### 2.1. Blockchain – how it works

Blockchain is a 40-year technology consisting of taking cryptographic hash, electronic signature, adding them up inside a block (information container), and chaining them together in a way that is permanent and unalterable. It is the underlying digital foundation that supports applications such as Bitcoin, Ethereum, Litecoin, etc. This emerging technology enhances the process of storing transactions and tracing assets in a cooperate network.

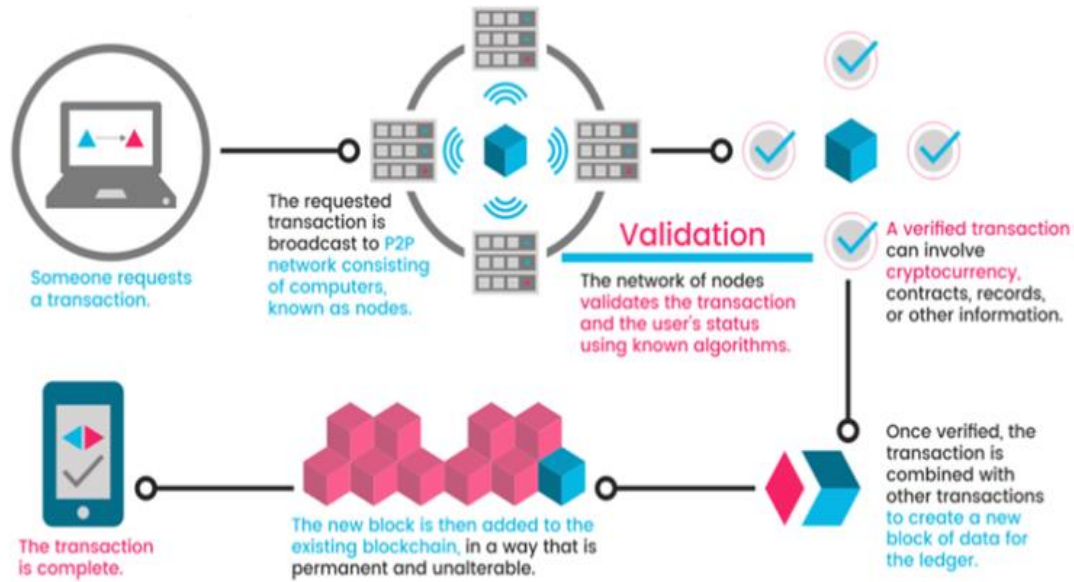

Figure 1. How blockchain technology works

## 2.2. Smart Contract

A conventional contract is an agreement between two or more parties that is enforceable by law; a smart contract is an agreement between two or more parties that lives on a blockchain and is enforceable by code.

## 2.3. Cryptocurrency

A cryptocurrency is a digital currency that relies on cryptography to secure the creation of new currency and transfer of funds, removing the need for a central issuing authority such as a central bank.

## 2.4. Distributed

All transactions can be seen on a distributed database; they can be validated by every partaking node contingent on the uniform rules [1]. Cryptography and consensus algorithms are fundamentally utilized to validate transactions, update the ledger, and maintain the core network security.

## 2.5. Auditability

In the distributed ledger, every transaction is "validated and recorded with a timestamp", this enables users to verify and trace the previous records from any node in the distributed network. It improves the traceability and the transparency of the data stored in the blockchain.

## 2.6. Immutability

Once a transaction has beencompleted in a blockchain network, the immutability of blockchain makes it nearly impossible to make changes, which increases confidence in data integrity and reduces opportunities for fraud.

## 3. CHALLENGES

Insurance industry encounters numerous challenges, and they keep growing over the time to become more consistent and complex. More recently, many insurance experts turn their attention to the potential of blockchain to address long-standing challenges related to the industry future and blockchain technology adoption. There are plenty of obstacles identified, but global regulations, trust and lack of accounting standards are eminent challenges for the world facing blockchain adoption [8].

## 3.1. (RE)Insurance Industry Challenges

Nowadays, when it comes to support, it's not a secret for anyone in service industries that customer needs and expectations are increasing and getting higher than ever before. High customer expectations lead to new challenges and create needs for new standards of services by insurance companies. The central finding of this document is that blockchain solutions have the potential to increase efficiency and improve outcomes, that can force the sector to find out ways to address the following challenges: (1) (re)insurers store fragmented records about assets, providers, and policyholders on various systems simultaneously, and the will of implementing blockchain based innovation is not an easy decision to pursue. (2) insurance applications fail to provide robust measures to secure participants information. (3) lack of visibility when information is sharing between insurers, peers, providers, claimants. (4) carriers are gathering pertinent information and documents from multiple sources. (5) multiple handoffs between systems contributors increasing time, cost, and risk of fraud.

## 3.2. Challenges for Blockchain Adoption

Like every technology, blockchain comes with multiple forms of challenges for its mass adoption. One of the major concerns in this adoption is in terms of security. Scalability of blockchain technology constitutes another major challenge for its adoption and implementation; transactions in blockchain system are validated through consensus mechanisms, the continuous replication, and the immutability of this technology lead to ever-growing amount of stored data. Industry professionals still want to see how well blockchain solution can perform in terms of integration with legacy systems to reduce total cost of its implementation. Other obstacle that needs to take into consideration is the lack of legal and regulatory framework since organizations have complex business rules and regulatory obligation to comply with.

## 3.3. Trust in Digital Economy

The key denominator of any economic or financial exchange is TRUST. Digital economy is an economy based on electronic goods and services and created by an electronic business. behind each transaction that is taking place requires the intervention of a trusted third party, even to claim ownership of an asset we rely on central authorities to verify and confirm our property rights [6].Since last decade, we have noticed a phenomenon called "Erosion of Trust" which is expressed as: distrust of central authority, the desire for freedom, the desire for privacy and anonymity, the distrust of intermediaries, and the distrust of corporations. First, and most obvious, are the fees that intermediaries charge for their services, which can be quite high. Secondly, relying on third party also implies cybersecurity risks, as retaining sensitive data on centralized servers creates a single point of failure for bad actors. Lastly, public confidence in financial establishments significantly deteriorated during the global financial crisis, and it may be more than mere coincidence that the Bitcoin protocol, which attempted to provide an alternative to the traditional financial system, was introduced in October 2008, as the global financial crisis was taking hold.

## 4. DRIVING FORCES AND OPPORTUNITIES

In recent years, multiple forces contribute to shape the future of the insurance market. Emerging technologies such as blockchain and artificial intelligence (AI), consumer expectations have become predominant drivers that have inflamed the competitive environment. According to Gartner.com, "by 2023 35% of enterprise blockchain applications will integrate with decentralized applications and services"[7]. The adoption of distributed ledger technologies by Payment networks along with decentralized finance (DeFi) applications are among of key drivers of the blockchain technology. The biggest concern remains the ability of blockchain implementation to eliminate the middlemen like brokers and agents sooner or later from the business process model of the insurance industry.

### 4.1. Emerging Technologies

The evolution of emerging technologies such as blockchain, artificial intelligence (AI), and machine learning offer a great opportunity for the current insurance system into a type of digital insurance platform. This approach will create new value, and new concept like "InsurTech" that stands for Insurance and Technology which is an insurance version of Fintech. The objective is to leverage technology power to increase productivity and efficiency in this market. To have an idea of how this technology could impact the global economy, on October 13th, 2020, London-UK, a report produced by PwC states Blockchain technology has the potential to boost global gross domestic product (GDP) by US$1.76 trillion over the next decade [5].

### 4.2. Customer Expectations

Many studies suggest that customer expectations are highest that it has never been before. Policy holders, beneficiaries, and other participants expect personalized and fast service at their finger. This new ongoing trend of social and technology are shaking in many ways the regular business patterns in the sector. Insurance agencies need to ensure customers their information remains private and have adequate measures in place to secure while sharing data with other entities. Prospects like options"and they are generally looking for advises and recommendations from the industry's Professionals. By establishing a consistent communication with their customers and assisting them in shopping process, are the effective ways for insurers to elevate customer service experiences.

### 4.3. Innovation and Disruption

Stephan Binder of McKinsey & Company characterizes the insurance industry as slow to change. He suggests that in the next 10 years, 40% of the insurance jobs that exists today will be gone, and 20% of the jobs in the future will be new. Conventional insurers and reinsurers are utilizing emerging technology to bring certain efficiencies into the industry which could potentially lead to higher financial gain or a more favorable competitive edge [10]. It may also result in a competitive drawback due to their long-established business models and their high-value investments in infrastructure. However, a growing number of newly formed insurers are focusing on consumers by offering better products and services with new business models designed to disrupt the traditional market. Despite of those pain points, new start-up carriers can take these opportunities to build a unique or different business model designed to exploit weaknesses in old-fashioned insurance's operations [9].

### 4.4. Insurance Market Initiatives and Ecosystem

Among Fintech initiatives, there are two major players capturing our attention: B3i - Launched in 2016, Blockchain Insurance Industry Initiative (https://b3i.tech/) which is an international industry blockchain technology consortium, its main purpose is to improve successes in data exchange among insurance and reinsurance enterprises [2].

RiskStreamCollaborative (https://web.theinstitutes.org/riskstream-collaborative) is the risk management and insurance industry's largest enterprise-level blockchain consortium that connects experts and developers to advance insurance specific use cases states its vision is to advance blockchain technologies and its capabilities to streamlining and bringing efficiencies in all areas of the risk management and insurance industries.

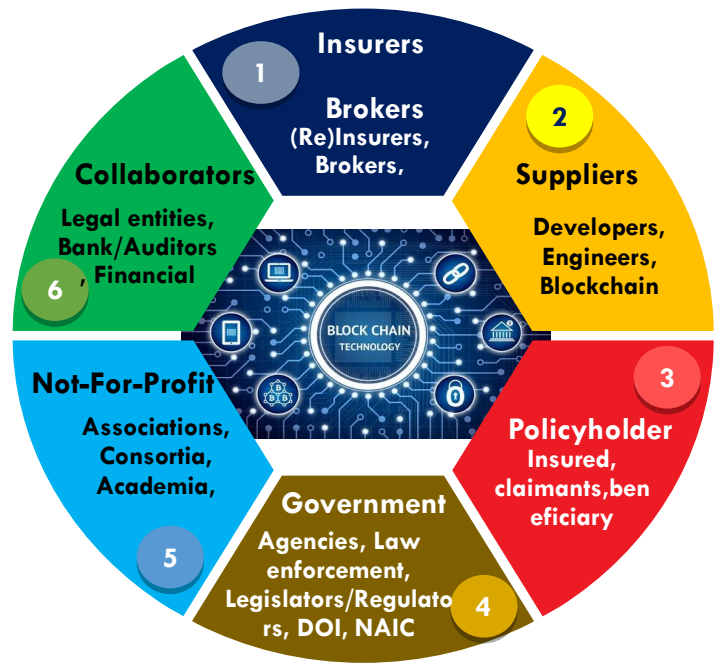

Figure 2. Blockchain insurance ecosystem

## 5. BLOCKCHAIN BUSINESS USE CASE

### 5.1. Insurance Claim Management

Majorly, there are two types of insurance product: life insurance and general insurance. We encounter two main categories of life insurance – permanent and term [3]. General insurance in other hand deals more with other form insurance related to valuables such as home, car, goods, and other hazard like fire. It is divided into parts namely motor insurance, health insurance,

International Journal of Network Security & Its Applications (IJNSA) Vol.14, No.6, November 2022

combined, comprehensive and package policies, property insurance, pecuniary insurance, and casualty insurance.

The following business use case is proving (re)insurance sector can leverage blockchain technology to streamline claims process, by reducing the industry average touch points from 26 to 7, cutting down the times for claims processing from 7-15 days to 72 hours, and lowering insurance premiums by 15% [4].

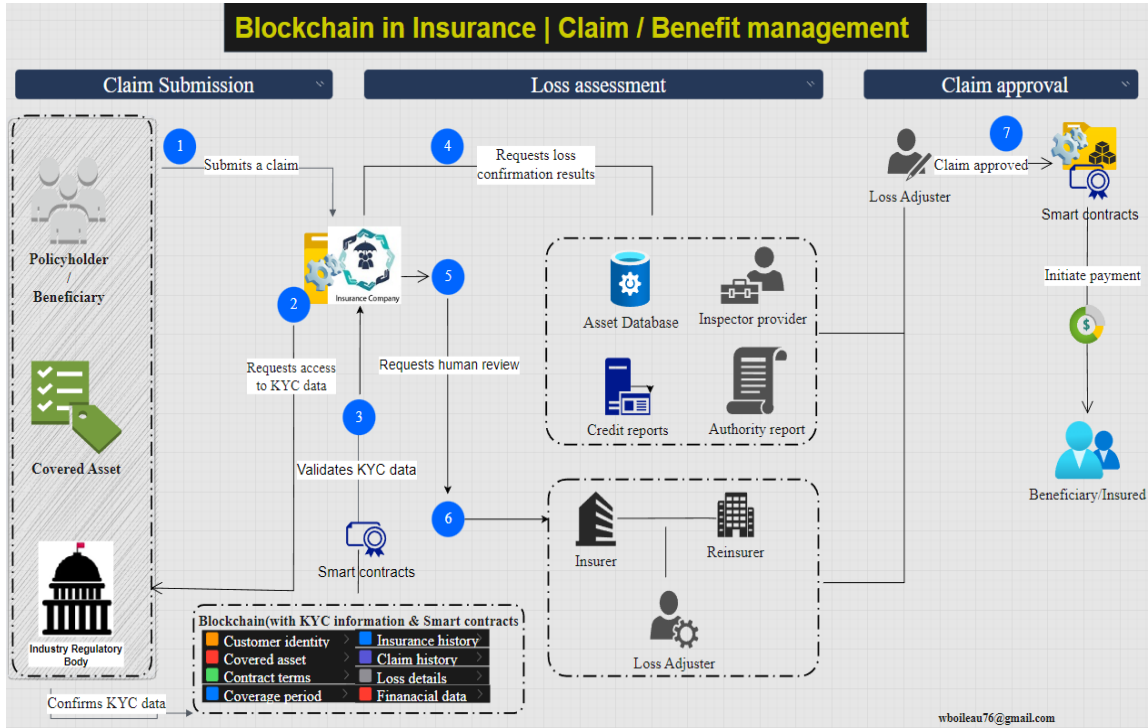

Figure 3. Claim/Benefit management use case

## 6. CONCLUSIONS

As has been noted, in section 3, we provide challenges for the insurance industry based on blockchain technology, section 5explains how blockchain can bring value and enable innovation in the value chain of the industry invariable giving rise to a new business model. As has been demonstrated, insurance industry can leverage blockchain technology to enrich crucial processes like claim submission and processing, anti-money laundering, fraud prevention. Firstly, while blockchain technology is making its baby steps, there are several promising use cases and applications for it in the insurance industry. We have already seen areas blockchain can lower operational cost and automate redundant process. Secondly, from the industry perspective, we discovered there is a great need for insurance companies to align around standards and processes within blockchain. While blockchain technology is developing better tools and framework for this market to collaborating and sharing data, insurance agencies themselves and other entities part of this ecosystem like regulators, financial institutions, and government must be willing to work with each other. The technology must go further in terms of development to address privacy and security, especially for public blockchain where everyone has access to each transaction in the ledger. Finally, since the insurance is highly regulated sector, legislators and regulators need to provide legal and regulatory frameworks, in addition laws and procedures currently in place for the technology to succeed.

International Journal of Network Security & Its Applications (IJNSA) Vol.14, No.6, November 2022

## REFERENCES

[1] Martino,Will – Kadena, the first scalable, high performance private blockchain, revision v1.0, August 2016

[2] Nicholson, E Jack, Cole, Cassandra and McCullough, Kathleen - Challenges for the Insurance Industry in the Future. National Association of Insurance Commissioners (NAIC), Journal of Insurance Regulation, Vol 38 No. 6, 2019.

[3] Akande, Adefisayo - Disruptive Power of Blockchain on the Insurance Industry. Tatu University Library 2018.

[4] Boileau,Wadnerson - Blockchain in insurance industry: turning threat into innovative opportunities, 2nd International conference on Cryptography and Blockchain, October 22-23 October 2022,Sydney, Australia.

[5] Blockchain technologies could boost the global economy US$1.76 trillion by 2030 (pwc.com)

[6] Michael Pisa and Matt Juden. 2017. "Blockchain and Economic Development: Hype vs. Reality." CGD Policy Paper. Washington, DC: Center for Global Development. https://www.cgdev.org/publication/blockchain-and-economic-development-hype-vs-reality

[7] Pwc, "Insurance 2020:Turning change into opportunity," *PwC Insur. 2020*, no. January, p. 24, 2012.

[8] https://blogs.gartner.com/avivah-litan/2021/07/14/hype-cycle-for-blockchain-2021-more-action-than-hype/

[9] R. Hans, H. Zuber, A. Rizk, and R. Steinmetz, "Blockchain and Smart Contracts: Disruptive Technologies for the Insurance Market," *Proc. Twenty-third Am. Conf. Inf. Syst. (AMCIS 2017)*, no. August, pp. 1–10, 2017.

[10] H. Kim, M. Mehar, Blockchain in Commercial Insurance: Achieving and Learning Towards Insurance That Keeps Pace in a Digitally Transformed Business Landscape, ‖ SSRN Electron J 2019.

## AUTHOR

**Wadnerson Boileau,** has a Master of Science in Business Analytics & Information Systems (2021) from University of South Florida (USF),and a bachelor's degree in Electronic Engineering from State University of Haiti(UEH). He started his career as Analyst Programmer/Developer where he built software for small businesses. In 2003, he became SAS Certified while implementing SAS Project for Haiti Central Bank. He joined an Oracle partner company providing support in processes and documentation. In 2014, after getting his Florida Licensed insurance, he worked with multiple Fortune U.S. 500companies as Financial Consultant. In 2019, he earned two Blockchain certifications.

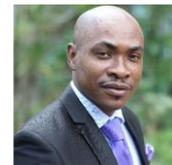